\documentclass[12pt]{iopart}

\usepackage{iopams}
\usepackage{graphicx}
\begin{document}

\title{Non-Markovian reservoir-dependent squeezing}   
\author{J. Paavola}
\address{
Turku Centre for Quantum Physics, Department of Physics and Astronomy, University of Turku,
FI-20014 Turun yliopisto, Finland
}
\ead{janika.paavola@utu.fi}

\date{\today}    

\begin{abstract}
The squeezing dynamics of a damped harmonic oscillator are studied for different types of environment without making the Markovian approximation. The squeezing dynamics of a coherent state depend on the reservoir spectrum in a unique way that can, in the weak coupling approximation, be analyzed analytically. Comparison of squeezing dynamics for Ohmic, sub-Ohmic and super-Ohmic environments is done showing a clear connection between the squeezing--non-squeezing oscillations and reservoir structure. Understanding the effects occurring due to structured reservoirs is important both from a purely theoretical point of view and in connection with evolving experimental techniques and future quantum computing applications.

\end{abstract}

\maketitle

\section{Introduction}\label{intro}

It is known from the theory of open quantum systems that the environment affects the state of the system by inducing decoherence and heating, which very quickly destroy quantum superpositions and entanglement \cite{Breuer}. This effect is a major obstacle in quantum information science which relies on the fragile quantum properties. Very often it is assumed that the environment does not have a memory, i.e., changes in the reservoir induced by the system do not have a back action effect on the system. This assumption is known as the Markovian approximation. However it is not valid in general for strong couplings or for certain reservoirs with long memory. These include, e.g., photonic band gap materials \cite{bandgap} and atom lasers \cite{atomlaser}. The need for non-Markovian theory depends on the properties of the environment. Recently, as experimental setups approach the limit where the Markovian approximation ceases to be valid, proposals for non-Markovian quantum computation have appeared in the literature \cite{Horodecki2002}.

When the non-Markovian effects are taken into account the evolution of the system is different from the Markovian case. To study non-Markovian effects I use the quantum Brownian motion model, describing a particle in a harmonic potential coupled to a quantized bosonic thermal reservoir \cite{Breuer}. An exact solution to the master equation describing the time evolution of the system exists \cite{Maniscalco2003,{HuPazZhang92}}. Of course the analytical form of the solution depends on the spectral distribution of the reservoir. In this paper I consider three different types of spectral distributions, Ohmic, super-Ohmic and sub-Ohmic. This allows to compare the effects of different reservoirs on a given quantum system which correspond to different physical realization of, e.g., a qubit. In this way we can clarify the microscopic processes underlying the dynamics of exemplary open quantum systems.

In this paper I consider an initially squeezed coherent state. The Heisenberg uncertainty principle for the variances of the dimensionless quadratures $x$ and $y$ states that $(\Delta x)^2(\Delta y)^2\ge1/4$. Squeezed states are the ones where the variance of one quadrature is smaller than that of the vacuum, i.e., $1/4$. The other quadrature will have a larger variance in order not to violate the uncertainty principle.
Previously we have studied the effects of an Ohmic, super-Ohmic and sub-Ohmic reservoir to the dissipation of a quantum harmonic oscillator \cite{Paavola2009} and the decoherence of a coherent superposition state \cite{Paavola2009b}. In this paper I investigate how the short-time non-Markovian dynamics of a squeezed state evolution is affected by these different reservoirs.

Squeezed states of light can be utilized in quantum information processing \cite{Furusawa08,Furusawa07} and quantum communication \cite{Yurke90} to improve error rates.
They have been used to construct entangled states and to demonstrate quantum teleportation of continuous variable quantum states \cite{Furusawa98,Bowen2003}. Squeezed states are also used in quantum metrology and high precision measurements \cite{Schnabel2007,Kimble87}. To the best of the author's knowledge, squeezed states of the harmonic oscillator have been studied in the non-Markovian regime only for the Ohmic reservoir \cite{Maniscalco2005}, where it was shown that the squeezing dynamics are a result of virtual processes between the system and the bath.

The paper is organized in the following way. In section II we present the model and master equation for the system. Section III introduces the different reservoir types used in the comparison. The main results of this paper are given in section IV. Finally section V contains the conclusion.
\section{The system and the master equation}
Our model is the quantum Brownian particle in harmonic potential which consists of a quantum harmonic oscillator linearly coupled to a bath of quantum harmonic oscillators. The Hamiltonian of the system and the bath, in units of $\hbar$, are $H_S=\omega_0 (a^\dagger a+1/2)$ and $H_E=\sum_n \omega_n (b_n^\dagger b_n+1/2)$, respectively, and the microscopic interaction Hamiltonian is
\begin{equation}
H_I=
\frac{g}{\sqrt{2}}(a+a^\dagger)\sum_n k_n
(b_n+b_n^\dagger),
\end{equation}
where $a (b)$ and $a^\dagger (b^\dagger)$ are the annihilation and creation operators of the system (bath), $g$ is a dimensionless coupling constant, $k_n$ gives the coupling between the system and each individual environment oscillator, and $\omega_0$ and $\omega_n$ are the frequencies of the system and the $n$th environment oscillator, respectively. The total Hamiltonian is then given as $H_{tot}=H_S+H_E+H_I$.
In the weak coupling limit (i.e., when $g\ll1$),
assuming an initially factorized state ($\rho=\rho_S\otimes\rho_E$) and
a thermal reservoir, we obtain the following secularly approximated
master equation for the damped harmonic oscillator \cite{Breuer,Messina2003}
\begin{eqnarray}\label{eq:mainME}
\frac{d}{dt}\rho_S(t)&=\frac{\Delta(t)-\gamma(t)}{2}\left(2a^\dagger
\rho_S\nonumber a-a a^\dagger\rho_S-\rho_S a
a^\dagger\right)\\
&+\frac{\Delta(t)+\gamma(t)}{2}\left(2a\rho_Sa^\dagger-a^\dagger
a\rho_S-\rho_Sa^\dagger a\right),
\end{eqnarray}
where
\begin{eqnarray}
\label{eq:delta}
\Delta(t) &= 2\int_0^{t}dt'\,
\int_0^{\infty}d\omega\,J(\omega)\left[N(\omega)+\frac{1}{2}\right]\\
&\times\cos(\omega\nonumber
t')\cos(\omega_0 t'), \\
\gamma(t) &= 2\int_0^{t}dt'\,
\int_0^{\infty}d\omega\,\frac{J(\omega)}{2}\sin(\omega
t')\sin(\omega_0 t'),\label{eq:pikkugamma}
\end{eqnarray}
are the diffusion and dissipation coefficients. $N(\omega) =
(e^{ \omega/k_B T}-1)^{-1}$ is  the average number of reservoir thermal excitations, with $k_B$  the Boltzmann constant and $T$ the
reservoir temperature, and $J(\omega)$ is the
spectral density of the environment defined, in the continuum limit, as
\begin{equation}
J(\omega)=\alpha^2\sum_n\frac{k_n^2}{m_n\omega_n}\delta(\omega-\omega_n),
\end{equation}
with $m_n$ the masses of the environmental oscillators. No Markovian approximation was done in obtaining equation (\ref{eq:mainME}) so this equation describes accurately the environment memory. The memory effects of the reservoir are contained in the time-dependent coefficients, given by equations (\ref{eq:delta}) and (\ref{eq:pikkugamma}). It has been shown that performing the secular approximation does not affect considerably the system dynamics, provided that we are in high $T$ weak coupling limit and restrict our attention to the parameter regime $r=\omega_c/\omega_0\ll 1$ \cite{Maniscalco09}, where $\omega_c$ is a reservoir parameter introduced in Section \ref{Sec:reservoir}. Now we can proceed to the description of the reservoirs.

\section{Modeling the reservoirs}\label{Sec:reservoir}

The spectral densities I examine are of the form
\begin{equation}\label{spectra}
J(\omega)=\alpha^2\omega_{c}^{1-s}\omega^s e^{-\omega/\omega_c}.
\end{equation}
The exponential cutoff with cutoff frequency $\omega_c$ is introduced to eliminate divergencies in the $\omega \rightarrow \infty$ limit.
The parameter $s$ appearing in equation (\ref{spectra}) is a constant that can acquire values $<1$, $1$ or $>1$, corresponding to the so called sub-Ohmic, Ohmic and super-Ohmic spectral densities, respectively. In this paper I consider some examples and fix the value of $s$ to $1/2$, $1$ and $3$. The three cases describe different physical contexts (see discussion in \cite{Paavola2009}).

The spectral distribution gives full information on the reservoir properties and it is given by $I(\omega)=J(\omega)\left[N(\omega)+\frac{1}{2}\right]$. In high $T$ this approximates to $I(\omega)=J(\omega)k_BT/\omega$. The parameter $r=\omega_c/\omega_0$ characterizes the overlap of the system frequency with respect to the reservoir.
\begin{figure}\label{fig:reservoirs}
\includegraphics[width=7cm]{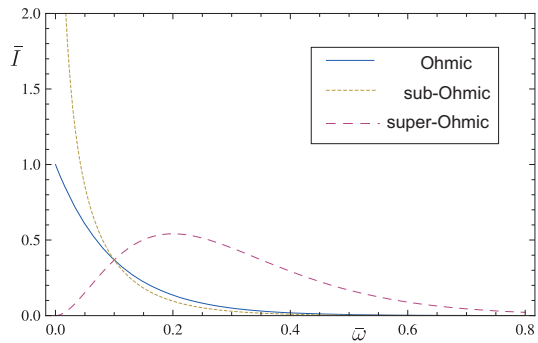}
\caption{Scaled spectral distribution $\bar{I}=I/(g^2k_BT)$ as a function of $\bar{\omega}=\omega/\omega_0$ for $r=\omega_c/\omega_0=0.1$.}
\end{figure}
%
In this paper I focus on a parameter region where $r\ll\ 1$. From previous studies \cite{Paavola2009,Maniscalco2004} we know that this is the regime where non-Markovian effects are most pronounced.  In figure 1 all three reservoirs are shown for $r=0.1$. With this choice of the parameters the time scale of the system dynamics, given by $\tau_S=1/\omega_0$, is much shorter than the relevant reservoir time scales $\tau_R=1/\omega_c$. For this reason one would expect the non-Markovian effects to be strong.
\section{Wigner function dynamics for an initial squeezed state}

The Wigner function is a phase space representation of the density matrix of the system and is thus a complete description of the state of the system. The solution of the master equation (\ref{eq:mainME}), based on algebraic properties of superoperators presented in \cite{Maniscalco2003}, is used here because it provides a solution in terms of the  quantum characteristic function
\begin{equation}\label{eq:chi}
\chi_t(\xi)=\mathrm{e}^{-\Delta_\Gamma(t)|\xi|^2}\chi_0[\mathrm{e}^{\Gamma(t)/2}\mathrm{e}^{-i\omega_0t}\xi],
\end{equation}
which can then be directly used to obtain the Wigner function. Here $\chi_0$ is the initial quantum characteristic function, $\xi$ is a complex variable, and
\begin{eqnarray}
\Gamma (t)&=&2\int_0^t\gamma(t_1)\,dt_1,\\
\Delta_\Gamma (t)&=&e^{-\Gamma(t)}\int_0^te^{\Gamma(t_1)}\Delta(t_1)\,dt_1.\label{muu}
\end{eqnarray}
We want to study initial squeezed coherent states. These states are obtained from the vacuum by operating first with the displacement operator $D=\mathrm{exp}(\alpha_0 a^\dagger-\alpha_0^*a)$ to obtain a coherent state, and then with the squeezing operator $S=\mathrm{exp}(1/2(za^2-z^*{a^\dagger}^2)$,
\begin{equation}
|\alpha_0,z\rangle=\hat{S}(z)\hat{D}(\alpha_0)|0\rangle.
\end{equation}
Here $z=se^{-i\phi}$ is the squeezing parameter.

The quantum characteristic function for an initial squeezed coherent state is
\begin{equation}
\chi_0(\xi)=\mathrm{exp}[-\frac{1}{2}|\xi C_s-\xi^*e^{-i\phi}S_s|^2+i(\xi^*\alpha_0^*+\xi\alpha_0)],\label{eq:char}
\end{equation}
where $C_s=\mathrm{cosh}(s)$ and $S_s=\mathrm{sinh}(s)$. By taking the Fourier transform of the quantum characteristic function (\ref{eq:chi}), with the help of equation (\ref{eq:char}), we obtain the Wigner function for the initially squeezed and displaced vacuum state ($\alpha_0=0$) with squeezing angle $\phi=0$
\begin{equation}
W_t(\alpha)=M\mathrm{exp}\left[\frac{-\alpha_x^2}{(\Delta x)^2(t)}+\frac{-\alpha_y^2}{(\Delta y)^2(t)}\right],
\end{equation}
where
\begin{eqnarray}\label{eq:variance}
(\Delta x)^2(t)&=&\Delta_\Gamma (t)+\frac{e^{-\Gamma(t)}e^{-2s}}{2}\\
(\Delta y)^2(t)&=&\Delta_\Gamma (t)+\frac{e^{-\Gamma(t)}e^{2s}}{2}
\end{eqnarray}
are the variances of the dimensionless quadratures $x=(a+a^\dagger)/\sqrt{2}$ and $y=-i(a-a^\dagger)\sqrt{2}$ and $M$ is a time-dependent normalization constant \cite{Maniscalco2005,Matsuo93}.

\begin{figure}\label{fig:varianssit}
\includegraphics[width=7cm]{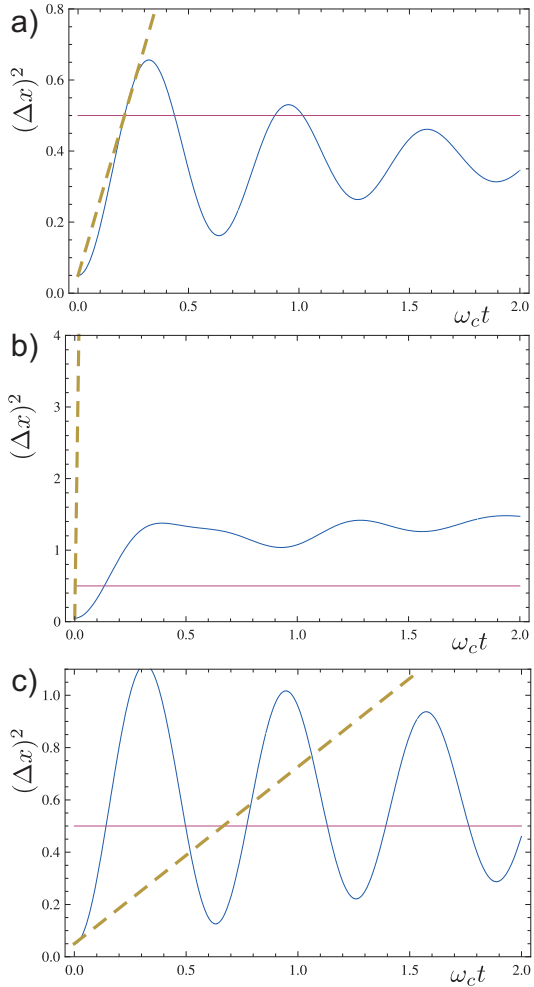}
\caption{Variance of quadrature $x$ for a) Ohmic, b) super-Ohmic and c) sub-Ohmic reservoir. Markovian result is given by dashed line. In b) the Markovian dynamics is so fast it appears here as an almost vertical line. The parameters are $g=0.1$, $\omega_c/\omega_0=0.1$, $\sigma^2=0.1$ and $k_B T/\omega_0=1.5\times 10^{3}$.}
\end{figure}
%
At the initial time $t=0$ we have $(\Delta x)^2(0)=e^{-2s}/2\equiv\sigma^2/2$ and $(\Delta y)^2(0)=e^{2s}/2\equiv 1/2\sigma^2$. In the following I consider the squeezing dynamics in the $x$ quadrature. The squeezing condition is $(\Delta x)^2<0.5$. For larger variances the state is not squeezed. An example of the squeezing dynamics for the Ohmic, super-Ohmic and sub-Ohmic reservoir is shown in figure 2.

Oscillations in the squeezing for non-Markovian times appear for Ohmic and sub-Ohmic reservoirs. They are induced by the temporarily negative values of the coefficients $\Delta(t)$ and $\gamma(t)$ which are known to occur at high $T$ and for $r\ll 1$, as in the case considered here. In this region, virtual exchanges of excitations between the system and the reservoir characterize the dynamics \cite{Maniscalco2004}. These virtual processes are causing the non-Markovian oscillations in the squeezing. The Markovian behaviour of the squeezing is plotted for comparison as a dashed line. The Markovian squeezing is obtained by inserting into equation (\ref{eq:variance}) the Markovian values of the coefficients $\Delta(t)$ and $\gamma(t)$, which are given as $\Delta_M=\pi J(\omega_0)$ and $\gamma_M=\pi I(\omega_0)/2$.

The parameters affecting the squeezing oscillations are the reservoir type, reservoir temperature and initial squeezing $\sigma^2$. The reservoir type is mostly responsible for the form of the curve, while the temperature and initial squeezing merely shift the curve with respect to the squeezing--non-squeezing border, $(\Delta x)^2=0.5$.

The variance $(\Delta x)^2$ exhibits very similar behaviour when coupled to Ohmic and sub-Ohmic reservoir but is different in character for the super-Ohmic case. The super-Ohmic reservoir induces non-monotonic dynamics in the variance, but for the choice of parameters used in the figures the squeezing oscillations do not appear. From the form of the curves we see that for all parameters, the super-Ohmic reservoir leads to a non-squeezed state in the shortest time.  Oscillations in the quadrature variance persist only as long as the decay rate $\Delta(t)$ attains negative values. For the super-Ohmic reservoir and $r\ll 1$ the decay rates attain positive values faster than the other two reservoirs \cite{Paavola2009}.

The differences between the squeezing for the super-Ohmic reservoir and for the Ohmic and sub-Ohmic reservoirs can be traced back to the reservoir properties. By comparing the figure 1 of the reservoirs with the squeezing plots in figure 2 we can easily come to the same conclusion as in \cite{Paavola2009} when considering the heating function. Namely, that the low frequency part of the spectrum $I(\omega)$ affects strongly the non-Markovian features. This means that when the spectrum is more tightly confined to the low frequencies (with respect to $\omega_0$), longer lasting oscillations both in squeezing and heating are present.

\section{Conclusions}
In this paper I have studied the non-Markovian squeezing dynamics of an initially squeezed coherent state. Analytic expressions for the variances of the quadratures depend on the reservoir structure. The effect of different reservoirs was studied for the system coupled to differently structured reservoirs. Depending on the reservoir type and other parameters, the initially squeezed state showed non-Markovian oscillations between squeezed and non-squeezed states.

The quantum Brownian motion model describes a large number of different physical systems \cite{applicationogQBM,einselection,Maniscalco04a,nuclear,chemistry}. Thus, understanding the short-time dynamics of this system contributes to a wide variety of experimental and theoretical scenarios.

Reservoir engineering techniques allow in principle to tune the reservoir parameters, especially the parameter $r$, in order to reach the oscillatory regime $r\ll 1$ here illustrated \cite{Wineland00}. The squeezing oscillations are connected to oscillations in the width of the Wigner function \cite{Maniscalco2005}, which can be measured with homodyne detection. Therefore the experimental detection of the non-Markovian squeezing dynamics is, in principle, possible.
The fact that non-Markovian dynamics are different for differently structured reservoirs may prove useful when considering implementing quantum devices with different physical setups, e.g. trapped ions or solid state materials. In any case understanding the effects of the environment on a quantum system is important also from a purely theoretical point of view.

\ack
 The author thanks Sabrina Maniscalco for helpful discussions and comments on the paper. Financial support from the V\"ais\"al\"a foundation is acknowledged.

\end{document}